\documentclass[sigconf]{acmart}
\usepackage{multirow}
\usepackage{booktabs}
\usepackage[most]{tcolorbox}
\usepackage{listings}
\usepackage{xcolor,xspace}
\usepackage{enumitem}
\usepackage{float}
\usepackage{subcaption}
\usepackage{dirtree}
\usepackage{xparse}      
\usepackage{hyperref}    
\usepackage{cleveref}
\usepackage{multicol}
\usepackage{makecell}

\crefname{questioncounter}{Question}{Questions}
\Crefname{questioncounter}{Question}{Questions}
\newcounter{questioncounter}[section]
\renewcommand{\thequestioncounter}{\thesection.\arabic{questioncounter}}

\NewDocumentEnvironment{questionbox}{o m} 
{%
\refstepcounter{questioncounter}%
\IfNoValueTF{#1}{\edef\question@lbl{question:\thequestioncounter}}{\edef\question@lbl{#1}}%
\begin{tcolorbox}[
    enhanced,
    breakable,
    colback=blue!5!white,
    colframe=blue!75!black,
    fonttitle=\bfseries,
    sharp corners,
    title={Question \thequestioncounter: #2}
]
\label{\question@lbl}%
}
{%
\end{tcolorbox}%
}

\definecolor{codegreen}{rgb}{0,0.6,0}
\definecolor{codegray}{rgb}{0.5,0.5,0.5}
\definecolor{codepurple}{rgb}{0.58,0,0.82}
\definecolor{backcolour}{rgb}{0.98,0.98,0.98}

\lstdefinestyle{pythonstyle}{
    backgroundcolor=\color{backcolour},
    commentstyle=\color{codegreen},
    keywordstyle=\color{magenta},
    numberstyle=\tiny\color{codegray},
    stringstyle=\color{codepurple},
    basicstyle=\ttfamily\footnotesize,
    breaklines=true,
    captionpos=b,
    keepspaces=true,
    numbers=left,
    numbersep=5pt,
    showstringspaces=false,
    tabsize=2,
    frame=single,
    rulecolor=\color{black!20},
    literate={...}{{\dots}}3
}
\lstnewenvironment{pythoncode}{\lstset{style=pythonstyle, language=Python}}{}

\newcommand{\PD}{Problem Decomposition\xspace}
\newcommand{\SRP}{Single Responsibility Principle\xspace}

\AtBeginDocument{%
  }

\setcopyright{acmlicensed}
\copyrightyear{2025}
\acmYear{2025}
\acmDOI{XXXXXXX.XXXXXXX}
\acmConference[ACE 2026]{ACM Conference}{February 2026}{Melbourne, Australia}%
\acmISBN{978-1-4503-XXXX-X/2025/06}

\begin{document}

\title{Assessing Problem Decomposition in CS1 for the GenAI Era}


\author[Samvrit]{Samvrit Srinath}
\orcid{1234-5678-9012}
\affiliation{
  \institution{University of California, San Diego}
  \city{La Jolla}
  \state{California}
  \country{USA}
}
\email{sasrinath@ucsd.edu}

\author[Annapurna]{Annapurna Vadaparty}
\orcid{0009-0002-4370-764X}
\affiliation{
  \institution{University of California, San Diego}
  \city{La Jolla}
  \state{California}
  \country{USA}
}
\email{avadaparty@ucsd.edu}

\author{David H. Smith IV}
\orcid{1234-5678-9012}
\affiliation{%
  \institution{Virginia Tech}
  \city{Blacksburg}
  \state{Virginia}
  \country{USA}
}
\email{dhsmith4@vt.edu}

\author{Leo Porter}
\orcid{1234-5678-9012}
\affiliation{%
  \institution{University of California, San Diego}
  \city{La Jolla}
  \state{California}
  \country{USA}
}
\email{leporter@ucsd.edu}

\author{Daniel Zingaro}
\orcid{0000-0002-1568-4826}
\affiliation{%
  \institution{University of Toronto Mississauga}
  \city{Toronto}
  \state{Ontario}
  \country{Canada}
}
\email{daniel.zingaro@utoronto.ca}






\renewcommand{\shortauthors}{Anonymous et al.}

\begin{abstract}

Problem decomposition---the ability to break down a large task into smaller, well-defined components---is a critical skill for effectively designing and creating large programs, but it is often not included in introductory computer science curricula. With the rise of generative AI (GenAI), students even at the introductory level are able to generate large quantities of code, and it is becoming increasingly important to equip them with the ability to decompose problems. There is not yet a consensus among educators on how to best teach and assess the skill of decomposition, particularly in introductory computing. This practitioner paper details the development of questions to assess the skill of problem decomposition, and impressions about how these questions were received by students. A challenge unique to problem decomposition questions is their necessarily lengthy context, and we detail our approach to addressing this problem using Question Suites: scaffolded sequences of questions that help students understand a question's context before attempting to decompose it. We then describe the use of open-ended drawing of decomposition diagrams as another form of assessment. We outline the learning objectives used to design our questions and describe how we addressed challenges encountered in early iterations. We present our decomposition assessment materials and reflections on them for educators who wish to teach problem decomposition to beginner programmers.  
\end{abstract}




\maketitle

\section{Introduction}
The proliferation of Generative AI (GenAI) tools like GitHub Copilot has caused a paradigm shift in computing education. While novices historically have spent large portions of their programming time writing code from scratch and struggling with syntax \cite{loksa2016programming}, a new reality has emerged: students can now generate correct code for almost any CS1 question with a single prompt~\cite{denny2023generativeai}. This capability has caused many to reconsider CS1 learning objectives ~\cite{cipriano2025programmers, vadaparty2024CS1LLM}. These educators are shifting focus away from some of the traditional learning objectives of CS1 courses that emphasize mastering language constructs, debugging syntax errors, and writing code from scratch \cite{robins2003learning}, focusing instead on skills that will remain critical 
as GenAI becomes increasingly responsible for the code-generation portion of software development. 

Problem decomposition--the ability to break complex tasks into well-defined, cohesive components--is one such skill. 
For computing education in general, problem decomposition is key for students to create high-quality code and demonstrate their understanding~\cite{haldeman2025teaching}. As students increasingly adopt GenAI tools for programming, problem decomposition becomes important for articulating requirements--what should each portion of their code do? Being able to clearly express the requirements of a problem and understand what the program is intended to do is critical for effective prompting ~\cite{10.1145/3731756}.  

Despite the importance of problem decomposition, it is not consistently taught across introductory computer science courses and is instead  often reserved for upper-division software engineering courses \cite{pearce2015improving}. In addition, existing pedagogical materials and assessment formats are not designed for a context where students have powerful code-generation tools available, leading some to consider course redesigns that include problem decomposition as a key learning objective \cite{cipriano2025programmers, vadaparty2024CS1LLM}. 


We similarly redesigned our CS1 course to focus on skills that have increasing relevance in a GenAI-integrated programming context, with a core design goal of explicit instruction on problem decomposition. 
We therefore integrate problem decomposition as a primary skill in our CS1 course, treating architectural thinking as a foundational competency alongside traditional programming skills such as code writing from scratch, code tracing, and code explaining. 

We developed new assessment formats that addressed the unique challenges of teaching decomposition. We created \textit{Question Suites}, multi-part scaffolded sequences of questions that guide students through a structured problem-solving process from conceptual understanding to architectural design. Each suite begins with a conceptual probe—a non-programming task ensuring that students understood the problem context—followed by component-level tasks (debugging, testing, interface design) that build familiarity with individual pieces, culminating in an architectural decomposition question where students select appropriate helper functions. 

In other assessments, we asked students to create \textit{Decomposition Diagrams} that required students to reason about program design in a purely visual, code-free format. Students created diagrams showing each function required for a program and how these functions related to each other, without needing to implement the functions. 

In this paper, we describe these new assessment strategies and our experience with implementing them throughout our redesigned CS1 course. As we deemed the approach promising, we offer this detailed account to help other educators navigate similar transitions. We report on the general format, the design principles that guided our development, preliminary data from students, and reflections on what worked and what proved challenging. 


\section{Related Work}\label{sec:related}
We consider what should be taught about decomposition in Section \ref{sec:related_principles}, 
and how people at varying levels of experience approach decomposition in Section \ref{sec:related_novice_experts}--this informs what we should teach and the challenges to keep in mind when teaching novices in particular. We then examine ways that prior instructors have taught decomposition in Section \ref{sec:related_teaching} and how that relates to the proliferation of GenAI in Section \ref{sec:related_genai}.

\subsection{Principles \& Best Practices of Decomposition}

\label{sec:related_principles}

The challenge of teaching problem decomposition is deeply connected to foundational principles from software engineering designed to manage complexity. A seminal contribution is Parnas's work on modularization, which argued for \textit{information hiding} as the primary criterion for decomposition \cite{parnas1972criteria}. This principle dictates that modules should hide difficult or likely-to-change design decisions, a strategy that prioritizes system changeability and results in modules with simpler, more abstract interfaces. Modularization also promotes reusability and reduces redundancy by encapsulating functionality that can be invoked from multiple contexts without duplication. This approach can also be viewed through the lens of Cognitive Load Theory; by encapsulating a complex decision, a module can be treated as a single conceptual "chunk" in working memory, reducing the extraneous cognitive load imposed on the developer.

Flowing from these ideas is the Single Responsibility Principle (SRP), which states that a module should have "only one reason to change" \cite{martin2003agile}. Adhering to SRP results in highly cohesive modules with fewer external dependencies (low coupling), making the system easier to understand and test. For a novice, SRP serves as a powerful and actionable heuristic, providing a concrete question they can ask of their code: "How many distinct responsibilities does this function have?"

\subsection{Decomposition by Novices and Experts}
\label{sec:related_novice_experts}
Prior work has consistently shown that experts dedicate significantly more cognitive effort to the initial stages of problem formulation and high-level planning than novices \cite{song2014expert, song2016problem}. Experts tend toward a breadth-first decomposition, where they first identify all major sub-problems at a high level of abstraction before exploring the implementation details of any single component \cite{jeffries2013processesinsoftware}. Novices, conversely, often exhibit a depth-first pattern, latching onto a single aspect of the problem and attempting to solve it completely before considering the larger system \cite{song2014expert}.

These strategic differences are rooted in the underlying organization of domain knowledge. Experts possess rich, interconnected knowledge structures, or schemas, allowing them to categorize problems based on deep principles \cite{elio1990modeling}. Novices, lacking these schemas, rely on surface-level features. Another common novice pattern is the Einstellung effect, or "mechanization of thought," in which a solver misapplies a known solution to a new, ill-suited problem \cite{EGIDI200615}. This novice-expert gap is a cross-domain phenomenon, observed in fields as varied as industrial design \cite{zhou2022analysis} and nursing \cite{sarsfield2014differences}.

\subsection{Teaching Practices for Problem Decomposition}
\label{sec:related_teaching}
Given the documented difficulties novices face, a significant body of research has focused on pedagogical strategies to teach decomposition explicitly. Early work by Sooriamurthi demonstrated the value of using a guided case study to guide novices through an incremental process of building and testing individual components before assembling them \cite{sooriamurthi2009introducing}. Pearce et al. found that a CS1 section receiving guided inquiry-based instruction on decomposition significantly outperformed a control group on final projects \cite{pearce2015improving}.

More recent work has analyzed the decomposition process itself. Charitsis et al. developed an automated system to analyze thousands of code snapshots from a CS1 course, finding that students who created functions earlier in the problem-solving process tended to complete assignments faster and score higher on the exam~\cite{charitsis2023detecting}. 

The challenge is fundamentally conceptual—decomposition difficulties arise at the design phase, independent of implementation details. This is evidenced by research using block-based programming, which shows that even when syntax barriers are removed, younger learners still struggle to decompose problems effectively as complexity grows \cite{kwon2019exploring}. Ultimately, effective decomposition is a metacognitive skill, part of a larger self-regulatory cycle of planning, monitoring, and reflecting on one's own problem-solving process \cite{loksa2016programming}.

\subsection{Decomposition in the GenAI Programming Landscape}
\label{sec:related_genai} 
The widespread availability of GenAI tools as code-generation assistants represents a paradigm shift for programming and, consequently, for computer science education. A major concern among educators is that students may use these tools to generate solutions without engaging in the underlying problem-solving process, leading to superficial learning \cite{lau2023ban}. However, this new landscape also reframes the primary learning objectives of CS1. As GenAI tools become proficient at translating a mental plan into correct syntax, the most critical human skill shifts from implementation to specification. A GenAI tool cannot generate a useful solution without a clear, well-structured prompt that accurately describes the desired components and their interactions, elevating the importance of decomposition to the central, assessable skill in AI-assisted programming \cite{cipriano2025programmers}. While prior work has established the importance of decomposition, documented the novice-expert gap, and begun to explore GenAI assistants as scaffolding tools, a critical gap remains in understanding how to teach and assess this skill in a CS1 course fully integrated with GenAI tools from the start. 

\section{Course Context}\label{sec:course_context}
Our introductory computing course was taught at a large, research-intensive university in North America over a 10-week term, followed by a final exam week. There were 181 students enrolled. The course met during three lecture times throughout each week, and there was one lab session each week in which students worked on an assignment with supervision from teaching staff.  The curriculum taught foundational programming concepts in Python, including variables, control flow, functions, strings, lists, and dictionaries, integrating the use of GitHub Copilot, a GenAI programming tool, from the first week. The latter half of the course explored applications in data processing, image processing, and game development.
Our course is modeled on that of \citeauthor{vadaparty2024CS1LLM}~\cite{vadaparty2024CS1LLM}.
The topics covered in the course are shown in Table ~\ref{tab:course_schedule}. We introduced \PD in Week 4. This timing ensured students first had a foundational grasp of Python code constructs and the purpose/use of functions before engaging with the higher-order cognitive task of structuring code. 

We introduced \PD through required readings from the course textbook ~\cite{porter2024learn}, followed by a professor-led, in-person lecture. Students were taught fundamentals such as the \SRP, which emphasizes that a function should have one primary responsibility. These ideas were subsequently reinforced through weekly homework assignments, where Question Suites dedicated to \PD were presented alongside other tasks like code writing, code tracing, explain in plain English (EiPE), testing, and debugging.

\begin{table}[ht]

\caption{Course Schedule}
\label{tab:course_schedule}
\begin{tabular}{|c|l|}
\hline
Week & Topic(s) \\
\hline \hline
1 & Functions and Working with Copilot\\ \hline
2 & Variables, Conditionals, Memory Models\\ \hline
3 & Loops, Strings, Testing, VSCode Debugger\\ \hline
4 & Loops, Lists, Files, Problem Decomposition \\ \hline
5 & Intro to Data Science, Dictionaries\\ \hline
6 & Revisit Problem Decomposition and Testing \\ \hline
7 & Intro to Images, PIL, Image Filters\\ \hline
8 & Copying Images, Intro to Games and Randomness\\ \hline
9 & Large Game Example\\ \hline
10 & Python Modules and Automating Tedious Tasks\\ \hline
\end{tabular}
\end{table}

\section{Question Suite Design}\label{sec:homework_design}
We describe our Question Suite methodology that we designed to scaffold and evaluate the high-level architectural thinking that has become a central  task in the age of GenAI-assisted programming. All of our materials are available at \url{https://anonymous.4open.science/r/problem_decomposition-4FD5/ProblemDecompositionCS1_compressed.pdf}.

\subsection{Learning Objectives and Approaches To Guide \PD Assessment Design}\label{subsec:guiding}
Our question development was guided by findings from prior work on problem decomposition. We now outline learning objectives for problem decomposition, and our approach toward designing questions that would help students reach these objectives. For the purpose of being able to assess and provide formative feedback in our large course setting, we chose to use multiple choice questions in our decomposition-related homework questions. 

\subsubsection{Learning Objective: Balancing Code Quality Trade-offs}
We drew inspiration from the conceptual framework proposed by Haldeman et al.\cite{haldeman2025teaching}, which defines optimal procedural decomposition as a balance of competing concerns. Our assessments were designed to compel students to reason about the inherent tensions that arise when structuring code. For instance, extracting duplicated code into a shared helper function eliminates redundancy. However, if that helper must accept many parameters to serve multiple calling contexts, it risks violating the \SRP by taking on multiple responsibilities. Additionally, requiring callers to pass information they don't naturally possess increases coupling. Conversely, strictly adhering to \SRP by creating highly specialized functions can necessitate passing the same data through multiple function calls, again increasing coupling. The challenge lies in finding decompositions that balance these concerns—minimizing duplication and maintaining cohesion without introducing unnecessary dependencies.

This tension aligns with Parnas's foundational goal of designing for change \cite{parnas1972criteria}. By making students explicitly consider these trade-offs—rather than mechanically applying a single rule—we encourage them to think architecturally about the long-term consequences of their design decisions. Our assessments therefore present scenarios where students must \textit{evaluate multiple decomposition strategies, each with different strengths and weaknesses}.

\subsubsection{Learning Objective: Problem-Solving Stages}
An important aspect of learning how to decompose problems is understanding that there are many stages required to do so. Our Question Suites were designed to guide students through a structured, reflective process, intended to counteract the documented novice tendency to rush into a depth-first implementation without a complete plan \cite{song2014expert}. We later found that our approach shared several key elements with the six-stage problem-solving model described by Loksa et al.\cite{loksa2016programming}. Specifically, our question suites prompted students to engage in similar stages, such as \textit{reinterpreting the problem prompt} before moving to \textit{implementation} and subsequent \textit{evaluation}.


\subsubsection{Learning Objective: Problem Solving in Novel Contexts}
A primary pedagogical goal was to foster student engagement by grounding \PD in authentic and motivating problem contexts. We recognized a key limitation of using only familiar scenarios: they do not adequately prepare students for a common reality in which programmers must rapidly learn and solve problems in completely new contexts.

Therefore, our methodology intentionally diverged from the approach presented in the course textbook. The textbook uses a familiar problem like a spell-checker, leveraging students' existing understanding of what a program is intended to do. Our Question Suites, in contrast, situate decomposition tasks within novel narratives—such as image processing algorithms, AI model pipelines, or game simulations—requiring students to first build an understanding of an unfamiliar domain before designing its architecture. This mirrors professional practice in which students will likely be faced with problem contexts that they will need to understand before creating solutions. Additionally, this focus on unfamiliar domains  works to prevent the Einstellung effect \cite{EGIDI200615}, where students might prematurely map a new problem onto a familiar but ill-suited solution pattern.

Note that we focus on this particular learning objective more in later homeworks--on earlier homeworks, we in fact leverage familiar problem contexts to avoid having students face the challenge of novel ones while still getting familiar with problem decomposition in the first place. 

\subsubsection{Learning Objective: Abstraction}
Our assessments deliberately concentrate on higher-level architectural decisions that require abstraction: identifying responsibilities, defining function interfaces, and structuring component hierarchies. This once again contrasts with the textbook, which covers top-down design but also includes detailed discussions of lower-level implementation, such as specific data structures for the spell-checker's dictionary. By separating the skill of decomposition from code comprehension and implementation—skills assessed elsewhere in our curriculum—we isolate and strengthen students' ability to think architecturally.

\subsection{Early Homeworks, Challenges, and Iterations} \label{sec:early_homeworks_challenges_iterations}
Our design of the \PD question suites evolved iteratively based on classroom observations and student performance. The central challenge was balancing complex, authentic problems with the need to manage cognitive load. Both the problem context and the required tasks grew in complexity over the term.

\subsubsection{First Decomposition Homework: Familiar Context and Impl\-ementation-Based Questions} For the first homework assignment that included decomposition questions (the homework assigned in Week 5 after students were first taught about decomposition in Week 4), our primary goal was to have students master the fundamental mechanics of decomposition. To achieve this, we intentionally limited the cognitive load by grounding questions in contexts we believed had a familiar and well-understood structure. An example of this first iteration situates the problem within the familiar setting of a course--the problem asks students to imagine that they are part of an instructional team analyzing student information, with data fields that students have likely already had to interact with in a data-entry context: their name and major. We intended for the pre-existing knowledge of anticipated inputs, outputs, and dataflow for a particular process to be a background that students were familiar with so that their focus could be on decomposing the problem rather than needing to spend a large quantity of time to understand it (similar to how the course textbook uses the example of a spellchecker). 
 
The tasks required in the first homework were also only implementation based ones, in contrast to the more abstract decomposition tasks that we aimed to build up to. For example, one question provided students with a large block of code that was not broken into functions, and asked students to identify best practices that it violated, rather than asking students to work with more abstract representations.


\subsubsection{Long Problem Descriptions May Lead to Guess and Check}
As we progressed to question suites with more complex and unfamiliar contexts, we encountered a significant pedagogical challenge: in order to define a novel context with enough specification to have a clearly correct solution in a multiple choice
 homework question, the description requires a significant amount of detail. 
This resulted in increasingly unwieldy problem descriptions.

The instructional team identified this as a critical flaw that may undermine our learning objectives. The sheer volume of text may have disincentivized reading and comprehending the entire problem. Instead of engaging with the context, we surmise that some students resorted to ``guessing and checking,'' cycling through answer options rapidly rather than engaging in sustained architectural reasoning. Conversations from teaching staff surfaced a potential rationale--it wasn't ``worth it" to the students to spend such a long time reading and understanding such a long problem description in order to answer a single question. 

\subsubsection{A Refined Approach: Conceptual Probes}
We needed to address this miscalibration, but wanted to continue asking novel questions that required rich context and therefore large quantities of time spent to read this context. We therefore decided to create more questions associated with each context, with the intention that students would find it worthwhile to read and understand the context if it would be necessary for more than just one question. 


We began each suite by front-loading the most critical context and leading with a \textbf{conceptual probe}. This initial question (or several questions) is a non-programming task that requires students to engage with the problem's domain logic. For instance, in one image tinting suite, we first ask students to perform a concrete numerical calculation based on the described approach. By asking students to demonstrate this baseline comprehension, we could be more confident that their subsequent work on the higher-level architectural decomposition tasks was a genuine reflection of their problem-solving ability, not a product of guesswork.

\subsection{Question Suite Structure}
Distilling what we learned from the first two homeworks that included problem decomposition (the homework assignments for weeks 5 and 6), we began to formalize the decomposition portions of the homework assignments into a consistent ``Question Suite'' structure. Each suite is not a single question, but rather a multi-part, narrative-driven experience designed to guide students through a holistic problem-solving process. This structure is a core component of our pedagogy, ensuring that students first comprehend the problem domain before tackling the high-level task of architectural design. While the exact components vary slightly to match the learning objectives of the week, the suites generally adhere to a core pedagogical pattern, which we detail below.

\subsubsection{The Conceptual Probe: A Comprehension Gate}
As mentioned above, we found it effective to begin each suite with a conceptual probe. This initial, non-programming task assesses a student's understanding of the problem's core logic and constraints, effectively acting as a "comprehension gate." For instance, in the image tinting suite, the initial question asks students to perform a specific numerical calculation based on the described algorithm. 
The primary purpose of these questions is to compel students to read and synthesize the context. However, no sequencing of homework questions was enforced--students were able to progress to subsequent parts of the suite even if they did not correctly answer the comprehension question.

\subsubsection{Component-Level Tasks: Fostering a Holistic View of the Problem}
Following the conceptual probe, each suite presents a series of two to three questions that focus on the individual components and behaviors of the larger system. The goal here is to broaden the student's perspective beyond low-level implementation and encourage them to consider the full lifecycle of a software component. This diverse set of tasks is a deliberate choice to reinforce the interconnectedness of design, testing, and debugging. We found it particularly valuable to include:
\begin{itemize}
    \item \textbf{Identifying Decomposition Flaws:} In earlier homeworks, we present students with a monolithic block of code and ask them to identify its design flaws, such as a lack of reusability or a violation of the Single Responsibility Principle. 
 This primes students to not only perform decomposition but also to recognize and articulate the characteristics of poor design.

    \item \textbf{Closed-Box Debugging:} These tasks describe a function's intended behavior alongside a summary of an observed bug, and students must diagnose the most likely logical error without seeing the code. For example, one question asks students to diagnose a color anomaly in a fade effect, a crucial skill that forces reasoning about a component's contract—its expected inputs and outputs—which is fundamental to integrating different pieces of a system.

    \item \textbf{Code Testing:} We also include questions that ask students to select a minimal but diverse set of test cases for a described helper function. In one suite involving a checkerboard image, students must select tests that cover the edges of the image, transitions across boundaries, and edge cases for square sizes, reinforcing the importance of considering testing conditions as an integral part of the design process.

    \item \textbf{Function Design:} 
     Another question type focuses on a function's interface. Students select the best-defined Python function header from several options. These questions heavily utilize \textbf{type hints and detailed docstrings} to shift the student's focus from implementation details to the function's contract and its specific role within the larger architecture. 
    
\end{itemize}
Conversely, we are intentional about the role of \textbf{targeted code-writing tasks}. While they are included for small, non-trivial helper functions—such as calculating a checkerboard square type—we use them sparingly. The primary goal of the suite is to develop architectural thinking, and we found that placing too much emphasis on implementation at this stage can distract from the main decomposition task. Code completion skills are assessed more thoroughly in other, dedicated sections of the homework assignments.

\subsubsection{The Function Selection Decomposition Question}
Each suite culminates in a question that assesses the primary learning objective: high-level problem decomposition. Having already engaged with the context and its components through the conceptual probe and component-level tasks, students are better prepared to reason about the overall architecture. In our Question Suite assessments (Homeworks 5 through 8), we employed a \textbf{function selection} format for this final task. 
In this format, students are presented with a list of potential helper functions and must select the subset that represents the most logical, effective, and appropriately granular decomposition for the main problem. This requires them to evaluate the cohesion and coupling of each potential function, as seen in the decomposition questions for the AI model and the circular color pop effect. This is a direct assessment of the design principles—such as the Single Responsibility Principle, cohesion, and coupling—addressed in Section~\ref{subsec:guiding}.

The function selection format is particularly well-suited for the scaffolded nature of Question Suites. By the time students reach this culminating question, they have already explored the problem domain through conceptual probes, engaged with individual components through debugging and testing tasks, and considered function interfaces through prototype selection questions. The decomposition question then synthesizes this understanding, asking students to identify which combination of helper functions best structures the solution.

\subsubsection{Example Question Suite: Circular Color Pop Effect}
To provide a concrete example of our Question Suite format, we present a condensed version of the ``Circular Color Pop'' suite from Homework 8. This suite exemplifies the scaffolded progression from conceptual understanding to architectural decomposition that characterizes our approach.

\paragraph{Conceptual Probe.} The suite begins by establishing domain understanding. Students are asked: \textit{``When converting a color pixel (R, G, B) to a grayscale pixel by averaging the R, G, and B values, which information is primarily lost and what is primarily preserved?''} This non-programming question ensures students understand that averaging preserves brightness (luminance) while losing hue and saturation—a prerequisite for reasoning about the effect's behavior.

\paragraph{Component-Level Tasks.} The suite then presents two intermediate questions:
\begin{itemize}
    \item \textbf{Closed-Box Debugging}: Students diagnose why a circular color pop function produces a colored circle in the top-left corner instead of the center. They must reason about coordinate systems and distance calculations without seeing code, identifying that the function likely uses \texttt{(0,0)} instead of the calculated image center when checking pixel distances.
    
    \item \textbf{Helper Function Evaluation}: Before the main decomposition question, students consider individual helper functions in isolation, such as \texttt{convert\_rgb\_to\_grayscale\_rgb()}, which transforms a single RGB tuple to its grayscale equivalent.
\end{itemize}

\paragraph{The Function Selection Decomposition Question.} Finally, students are presented with the full task specification and asked to select ALL appropriate helper functions for implementing \texttt{apply\_circular\-\_color\_pop}. The function must: (1) determine the image center and radius, (2) iterate through pixels, (3) check if each pixel is inside the circle, and (4) convert pixels outside the circle to grayscale. Students evaluate seven potential helper functions, shown in condensed form in \Cref{fig:circular_color_pop_suite}.

\begin{figure}[t]
    \centering
    \small
    \begin{tcolorbox}[title=Circular Color Pop: Decomposition Question (Condensed)]
        \textbf{Task:} Design \texttt{apply\_circular\_color\_pop(img, radius\_ratio)} to keep a circular region in color while converting the rest to grayscale.
        
        \vspace{0.5em}
        \textbf{Select ALL appropriate helper functions:}
        
        \vspace{0.5em}
        \begin{enumerate}[label=\Alph*), leftmargin=*, itemsep=2pt]
            \item \textbf{[Correct]} \texttt{calculate\_circle\_properties(width, height, ratio) -> tuple}: \\
            Computes center coordinates and radius in pixels.
            
            \item \textbf{[Correct]} \texttt{convert\_rgb\_to\_grayscale\_rgb(color) -> tuple}: \\
            Converts an RGB tuple to its grayscale equivalent.
            
            \item \textbf{[Correct]} \texttt{is\_inside\_circle(px, py, center\_x, center\_y, radius) -> bool}: \\
            Checks if a pixel falls within the circle's boundary.
            
            \item \texttt{process\_entire\_image\_for\_selective\_color(img, style) -> Image}: \\
            A high-level function applying various selective color styles. \textit{[Too generic]}
            
            \item \texttt{get\_image\_metadata(image) -> dict}: \\
            Extracts EXIF data from an image file. \textit{[Unrelated to task]}
            
            \item \texttt{apply\_grayscale\_to\_pixel(img, x, y) -> None}: \\
            Modifies a pixel in-place to grayscale. \textit{[Poor abstraction—returns None]}
            
            \item \texttt{copy\_image\_region(source, target, coords) -> None}: \\
            Copies a rectangular region between images. \textit{[Wrong approach]}
        \end{enumerate}
    \end{tcolorbox}
    \caption{Condensed version of the Circular Color Pop decomposition question. Students must identify which helper functions provide logical, appropriately-granular decomposition. Options A-C are correct; they encapsulate distinct computational steps (geometric setup, color transformation, and geometric condition checking).}
    \label{fig:circular_color_pop_suite}
\end{figure}

This example illustrates how the Question Suite structure guides students from conceptual understanding through component-level reasoning to architectural decomposition. The correct answers (options A, B, and C) each encapsulate a distinct, reusable computation. By the time students reach the decomposition question, they have already reasoned about these individual pieces, making the architectural synthesis more tractable.


\section{Decomposition Diagrams}\label{sec:diagrams}
While Question Suites represent our primary assessment strategy for problem decomposition, we also employed a complementary approach: diagram-based tasks that require students to reason about system architecture in a purely visual, code-free format. A \textit{decomposition diagram} shows all of the functions for a particular program, with lines to indicate which functions call other ones (examples are shown in Figure~\ref{fig:car_simulator_diagram}). When students reason about a system's high-level architecture in diagrammatic form, they must think about the system \textit{at the outset}, before engaging with implementation details or component behaviors. This ordering reflects the expert problem-solving strategy we aim to cultivate: breadth-first planning, where the overall system architecture is established before diving into the details of any single component \cite{song2014expert}. Diagrammatic tasks mirror authentic software design practice, where high-level architectural decisions must often be made before detailed specifications are available. Here we describe questions we ask students in which they must create or select correct decomposition diagrams in a lab exercise and in their homework. 



\subsection{The Evil Word Guesser Lab Task}

In their Week 9 lab activity, students were required to \textit{generate} a decomposition diagram from scratch given a problem description.

Students were tasked with creating a complete function decomposition diagram for the ``Evil Word Guesser'' game \cite{schwarz2011evilhangman}. The design of this problem intentionally deviates from the traditional version of the Word Guesser game to create a significant decomposition challenge. In the traditional Word Guesser, the computer's logic is straightforward: pick a single secret word and check user guesses against it. The program state is minimal, and the decomposition often maps directly to a basic game loop.

Our ``Evil'' variant introduces algorithmic complexity that demands careful architectural thinking. Instead of committing to a single word, the computer manages a dynamic list of all possible words. After each guess, it partitions the entire word list into ``word families'' based on the guessed letter's position, and then strategically selects the largest family to maintain maximum ambiguity. This deviation from the standard game forces students to consider a more sophisticated architecture.

Students must conceptualize and create distinct functions for: (1) managing and filtering the word list based on word length, (2) generating the word family patterns for a given letter and word list (e.g., ``- - a - a'' would be one of the word family patterns that would be generated after the guess `a', and would include words like ``koala''), (3) partitioning the word list into families, and (4) determining the optimal family to choose based on size. These are tasks absent in the simpler version. The challenge represents what Parnas describes as a ``difficult design decision'' \cite{parnas1972criteria}---the word list management and family selection logic must be hidden from the main game loop through appropriate functional decomposition.

In their lab, students were asked to read the problem description and then draw a decomposition diagram on a blank sheet of paper.  This open-ended format provides an authentic assessment of students' ability to apply top-down design to a novel problem without the scaffolding of pre-defined options. 


\subsection{Patterns in Diagrammatic Decomposition Lab Task} 
Many students correctly completed the task, and teaching staff did not notice any persistent confusions or difficulty (though they were instructed to only answer clarifying questions and not to help students with solving the problem). 

Some students used the diagrams to illustrate data flow rather than functions; that is, the boxes and descriptions in their diagrams represented the data rather than functions (and some students would show both), in contrast to the instructions which stated that their diagrams should show each function along with a description of what the function does. This highlights an interesting limitation of decomposition diagrams: a diagram that shows only functions and and arrows to indicate which ones call each other \textit{does not} have any way to represent data flow. Variables created inside of a main function, for example, might be necessary to think about in order to decide how to decompose the problem, but wouldn't explicitly be represented in a functions-only diagram. 

Another interesting pattern is that some students used arrows to show which functions called each other as we intended, while other students used arrows to show sequencing, with arrows going from functions that were called first to the ones called after them. This similarly illustrates information that would not be shown in our intended decomposition diagrams--they show which functions call each other, but do not show the order in which different steps occur.

\subsection{Diagrammatic Decomposition in Homework}

We included diagrammatic decomposition questions in the final homework assignment (Homework 9). These questions present students with four hierarchical diagrams representing alternative decompositions of a complex system, and ask them to select the diagram that best represents a balanced, maintainable top-down design. This format was also included in the final exam, allowing students to demonstrate their diagrammatic reasoning abilities in a summative setting after having received practice with it in the homework. 

\subsubsection{Example: Turn-Based Car Simulator}

\begin{figure*}[t]
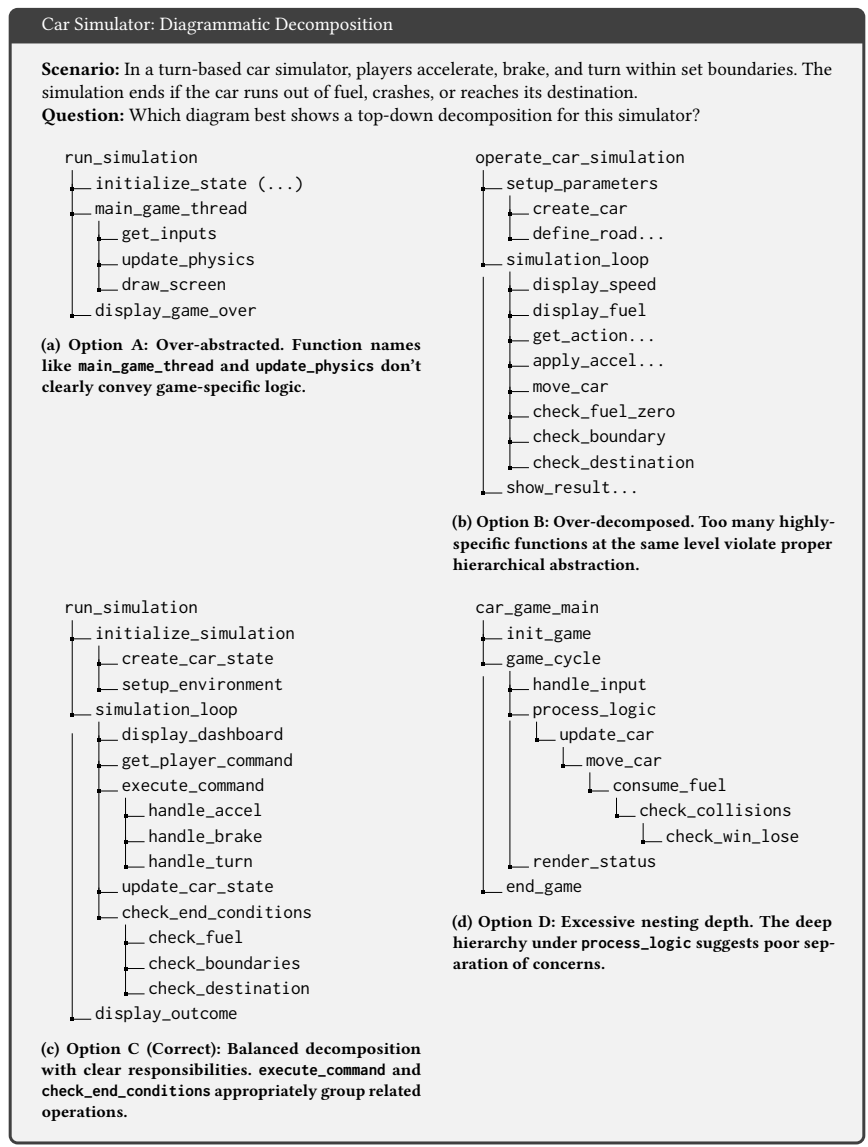

    \centering
    \scalebox{0.8}{
    \begin{minipage}{0.8\linewidth}
        \begin{tcolorbox}[title=Car Simulator: Diagrammatic Decomposition]
            \textbf{Scenario:} In a turn-based car simulator, players accelerate, brake, and turn within set boundaries. The simulation ends if the car runs out of fuel, crashes, or reaches its destination.
            
            \textbf{Question:} Which diagram best shows a top-down decomposition for this simulator?
            
            \vspace{1em}
            \begin{subfigure}[t]{0.48\linewidth}
                \dirtree{%
                .1 {run\_simulation}.
                .2 {initialize\_state (...) }.
                .2 {main\_game\_thread}.
                .3 {get\_inputs}.
                .3 {update\_physics}.
                .3 {draw\_screen}.
                .2 {display\_game\_over}.
                }
                \caption{Option A: Over-abstracted. Function names like \texttt{main\_game\_thread} and \texttt{update\_physics} don't clearly convey game-specific logic.}
            \end{subfigure}
            \hfill
            \begin{subfigure}[t]{0.48\linewidth}
                \dirtree{%
                .1 {operate\_car\_simulation}.
                .2 {setup\_parameters}.
                .3 {create\_car}.
                .3 {define\_road...}.
                .2 {simulation\_loop}.
                .3 {display\_speed}.
                .3 {display\_fuel}.
                .3 {get\_action...}.
                .3 {apply\_accel...}.
                .3 {move\_car}.
                .3 {check\_fuel\_zero}.
                .3 {check\_boundary}.
                .3 {check\_destination}.
                .2 {show\_result...}.
                }
                \caption{Option B: Over-decomposed. Too many highly-specific functions at the same level violate proper hierarchical abstraction.}
            \end{subfigure}

            \vspace{1em}
            
            \begin{subfigure}[t]{0.48\linewidth}
                \dirtree{%
                .1 {run\_simulation}.
                .2 {initialize\_simulation}.
                .3 {create\_car\_state}.
                .3 {setup\_environment}.
                .2 {simulation\_loop}.
                .3 {display\_dashboard}.
                .3 {get\_player\_command}.
                .3 {execute\_command}.
                .4 {handle\_accel}.
                .4 {handle\_brake}.
                .4 {handle\_turn}.
                .3 {update\_car\_state}.
                .3 {check\_end\_conditions}.
                .4 {check\_fuel}.
                .4 {check\_boundaries}.
                .4 {check\_destination}.
                .2 {display\_outcome}.
                }
                \caption{\textbf{Option C (Correct)}: Balanced decomposition with clear responsibilities. \texttt{execute\_command} and \texttt{check\_end\_conditions} appropriately group related operations.}
            \end{subfigure}
            \hfill
            \begin{subfigure}[t]{0.48\linewidth}
                \dirtree{%
                .1 {car\_game\_main}.
                .2 {init\_game}.
                .2 {game\_cycle}.
                .3 {handle\_input}.
                .3 {process\_logic}.
                .4 {update\_car}.
                .5 {move\_car}.
                .6 {consume\_fuel}.
                .7 {check\_collisions}.
                .8 {check\_win\_lose}.
                .3 {render\_status}.
                .2 {end\_game}.
                }
                \caption{Option D: Excessive nesting depth. The deep hierarchy under \texttt{process\_logic} suggests poor separation of concerns.}
            \end{subfigure}
        \end{tcolorbox}
    \end{minipage}
    }
    \caption{Diagrammatic decomposition question for the car simulator. Students must evaluate four architectural approaches, each representing different design trade-offs. Option C provides the best balance: clear two-level hierarchies under \texttt{simulation\_loop}, with \texttt{execute\_command} appropriately grouping the three player actions and \texttt{check\_end\_conditions} grouping the three termination checks.}
    \label{fig:car_simulator_diagram}
\end{figure*}

To illustrate this format, \Cref{fig:car_simulator_diagram} presents the car simulator decomposition question. Students are given a brief scenario---a turn-based driving simulation with fuel management, boundary checking, and destination reaching---and must evaluate four alternative architectural approaches. Each option represents a different balance of granularity, cohesion, and coupling.

The correct answer, Option C, demonstrates several key principles. First, it maintains \textit{consistent levels of abstraction}: the main \texttt{simulation\_loop} delegates to functions like \texttt{get\_player\_command} and \texttt{update\_car\_state}, each of which has a clear, singular responsibility. Second, it appropriately \textit{groups related operations}: the three player actions (accelerate, brake, turn) are encapsulated under \texttt{execute\_command}, while the three termination conditions are grouped under \texttt{check\_end\_conditions}. This grouping reduces coupling by hiding implementation details from the main loop. Third, the decomposition is \textit{neither too shallow nor too deep}---contrast with Option A's generic abstractions or Option D's excessive nesting.

Following this diagrammatic question, students are asked to define function interfaces for several of the components shown in their selected diagram. For example, they must choose the best prototype and docstring for \texttt{update\_car\_state}, evaluating options that differ in parameter granularity (individual values vs. dictionaries), type hints (specific vs. vague), and return strategies (new value vs. in-place modification). These follow-up questions assess whether students can translate their high-level architectural understanding into concrete design decisions about data flow and component contracts.


\section{Future Work}\label{sec:future_work}
To better incentivize careful problem analysis, future iterations of courses aiming to teach decomposition could implement \textit{required} conceptual probes, preventing progression to the next part of the question suite until comprehension is demonstrated. 

Future work could also explore alternative assessment formats that more directly measure architectural reasoning. One promising direction is refactoring tasks, where students receive a non-decomposed function and must decompose it into helper functions while preserving behavior. Such tasks would help students appreciate technical debt and practice an authentically professional skill. Parson's Problems---reordering pre-written code fragments---could also serve as an effective intermediate format between selecting a high-level design and writing code from scratch, potentially reducing cognitive load while maintaining rigorous assessment of structural understanding.

Measuring the full breadth of students' decomposition skills remains an open challenge. For example, we likely need to move beyond multiple choice questions and decomposition diagrams if we aim for students to decompose and implement problems on their own. Developing more automated, objective metrics for evaluating decomposition diagrams would facilitate scaling these assessments to larger populations and multiple institutions. This would enable the community to better understand whether our pedagogical approach transfers to different course contexts and student populations.

Future work could also consider ways to include additional elements not currently captured in decomposition diagrams. As shown by some students' inclination to illustrate sequencing and the flow of data in their diagrams, there is information important to program design that function-only decomposition diagrams do not capture. Possible ideas for future instruction could include ways to represent aspects like data and sequencing so that diagrams can be more expressive and better coupled to the designer's intentions.

\section{Conclusion}\label{sec:conclusion}
As AI tools continue to automate code implementation, the ability to analyze, plan, and decompose novel problems remains the essential human skill. 
We need to prepare students for this new reality. We see the current moment as critical for CS1 as we transition from pre-GenAI to post-GenAI courses. To maintain its relevance, we argue that we need to integrate problem decomposition into CS1, rather than leaving this topic for upper-year courses.

In this paper, we presented our experience of assessing problem decomposition in a GenAI-integrated CS1 course. We introduced the ``Question Suite,'' a scaffolded assessment methodology designed to guide novices through a structured problem-solving process that mirrors expert practice: understanding the domain, exploring individual components, and synthesizing understanding into architectural decisions. We complemented this with diagrammatic decomposition tasks---both formative (through homework) and summative (through lab and exam)---that isolate architectural reasoning in a code-free format. We offer our question suites to the community as examples of how problem decomposition can be comprehensively assessed.



\begin{acks}
Anonymized for review
\end{acks}

\bibliographystyle{ACM-Reference-Format}
\bibliography{refs}

\end{document}